# Gate Electrostatic Controllability Enhancement in Nanotube Gate all Around Field Effect Transistor

Laixiang Qin[1], Chunlai Li[2], Ziang Xie[3], Yiqun Wei[2], Jin He[1*]

1. Shenzhen SoC Key Laboratory, PKU-HKUST Shenzhen-Hong Kong Institution, Shenzhen 518057, China

2. Shenzhen institute of Peking University, Shenzhen 518057, China

3. Beijing Key Laboratory for Magneto-Photoelectrical Composite and Interface Science, School of Mathematics and Physics, University of Science and Technology Beijing, Beijing 100083, People's Republic of China

* Electronic email: frankhe@pku.edu.cn

**Abstract**

   Recently, short channel effects (SCE) and power consumption dissipation problems pose big challenges which need imperative actions to be taken to deal with for field effect transistor to further scale down as semiconductor technology enters into sub-10nm technology node. From 3nm technology node and beyond, gate all around field effect transistor steps onto the history stage attributed to its improved SCE suppressing ability thanks to surrounding gate structure. Herein, we demonstrate the super electrostatic control ability of a double-gated nanotube gate all around field effect transistor (DG NT GAAFET) in comparison with nanotube (NT GAAFET) and nanowire gate all around field effect transistor (NW GAAFET) with the same device parameters designed. $I_{on}$ boosts of 62% and 57% have been obtained in DG NT GAAFET in comparison with those of NT GAAFET and NW GAAFET. Besides, substantially suppressed SCEs have been obtained in DG NT GAAFET due to enhanced electrostatic control, which are certificated by improved $I_{off}$, SS, and $I_{on}/I_{off}$ ratio obtained. On the other hand, the $I_{on}$ of NT GAAFET is comparable with that of NW GAA-FET. Whereas, its $I_{off}$ is 1 order smaller, and SS is almost 2 times smaller compared with those of NW GAA-FET, manifesting the meliority of nanotube channel structure. In the end, the robustness of nanotube channel structure, especially double gated one, against $L_g$ scaling has been verified with TCAD simulation study.

**Introduction**

   As field effect transistor keeps on downscaling to sub-10nm node, SCE degradation and power dissipation increment grow increasingly unbearable these days [1-3]. SCE that degrades device performance seriously includes subthreshold swing (SS), which describes the smallest gate voltage needed to increase the drain current, $I_{ds}$, by a decade at subthreshold region; Drain induced barrier lower, (DIBL), which denotes the effect that the drain voltage, $V_{ds}$, will compete for the control ability of channel over gate; Leakage current increase, which contributes to power consumption increase; and subthreshold voltage decrease and so forth. Fortunately, SCE can be suppressed efficiently when the channel length ($L_g$) is 3 to 5 times larger than the screen length, $\lambda$, according to the former studies [4-5]. $\lambda$ can be expressed as,

$$\lambda = \sqrt{\frac{\varepsilon_{body}}{N\varepsilon_{ox}} t_{ox} d_{body}} \qquad (1)$$

$\lambda$ represents the channel region that is electrostatically controlled by the drain. In the band diagram, $\lambda$ is on behalf of the bending areas between the source to channel and channel to drain. Obviously, to enhance the inhibition ability over SCE of a transistor, small $\lambda$ is preferred. To decrease $\lambda$, aside from decreasing $t_{ox}$, $d_{body}$, $\varepsilon_{body}$ and increasing $\varepsilon_{ox}$ by using high k oxide, another effective method is to increase gate numbers, N, as expressed in equation (1). Fin field effect transistor (FinFET), with three sides of the channel surrounded by the gates, proved to be superior in suppressing SCE than planar FET [6-8]. From 22nm to 5nm technology nodes, FinFET has performed effectively, whereas, when transistor further scales to 3nm node and beyond, FinFET structure begins to be confronted with challenges of increasingly large leakage current, which both results from worse SCE and increasing tunnel leakage between source and drain regions [9]. Under these circumstances, GAAFET begin to draw researchers' attention due to its superior gate electrostatic control ability by virtue of the surrounding gate structure [10-13]. GAAFET lives up to the researchers' expectations and promises to sustain Moore's law to 0.7nm node according to simulation study performed by Yanling Shi et al [14]. Now that aggrandizing the gate number N is useful in suppressing SCE, nanotube GAAFET with both inner and outer channels surrounded by inner and outer gates simultaneously is predicted to perform better than nanowire GAAFET. There have been several proofs performed formerly [15-19]. Whereas, barely no studies notice the excellent scalability of nanotube channel structure apart from the super electrostatic control of both the inner and outer gates.

Herein, we launch a simulation study of double-gated nanotube GAAFET with both inner and outer channels surrounded by inner and outer gates concurrently to shed light on the superior electrostatic control ability of the DG structure and outstanding scalability of nanotube channel structure. The inner and outer gates are of the same length. For comparison, we also simulate a nanotube GAAFET with just a outer gate and a nanowire GAAFET as well. All three GAAFET structures are with the same device parameters, only that the DG NT GAAFET has an additional inner gate in the middle instead of a hole as in the NT GAAFET structure. The DG NT GAAFET obtain a 62% and 57% $I_{on}$ gains compared with those of NT GAAFET and NW GAAFET. Typical SCE figure of merits as $I_{off}$, SS and $I_{on}/I_{off}$ ratio have been ameliorated substantially, indicative of dramatically suppressed SCE attributed to enhanced gate electrostatic controllability in DG NT GAAFET. For another, the $I_{on}$ of NT GAAFET is comparable with that of NW GAA-FET. Whereas, $I_{off}$ and SS are enormously improved compared with NW GAA-FET due to thinner channel thickness leading to enhanced electrostatic control. Besides, the scaling behaviors of the three kinds of GAAFETs have also been accessed with TCAD simulation. According to the simulation results, DG NT GAAFET shows a better scaling behavior than both the NT and NW GAAFET structures, and meanwhile, NT GAAFET shows superior scalability than NW GAAFET when $L_g$ scales down, indicating that the nanotube channel structure possesses superior scalability than nanowire channel structure both in suppressing SCE and alleviating power dissipation problems.

**Device structure and Simulation Setups**

The 3-D structures of DG NT GAAFET, NT GAAFET and NW GAAFET and their corresponding channel profiles perpendicular to and along the y axis simulated in this study are shown in Fig. 1. The gate lengths selected for all three GAAFET structures are 50nm, the source/drain extensions length are halves of $L_g$. Nanotube thickness of DG NT GAAFET and NT GAAFET is 10nm, inner core radius is 20nm, nanowire radius of NW GAAFET is 30nm. Source/drain have doping concentrations of $1\times20/cm^3$ with phosphorus. The channels are doped with boron with concentrations of $1\times17/cm^3$. The high k oxide is $HfO_2$, with a thickness of 2nm. The models used are drift diffusion, high field saturation velocity, doping degradation, Shockley-Read-Hall (SRH) Generation Recombination, Bandgap Narrowing models.

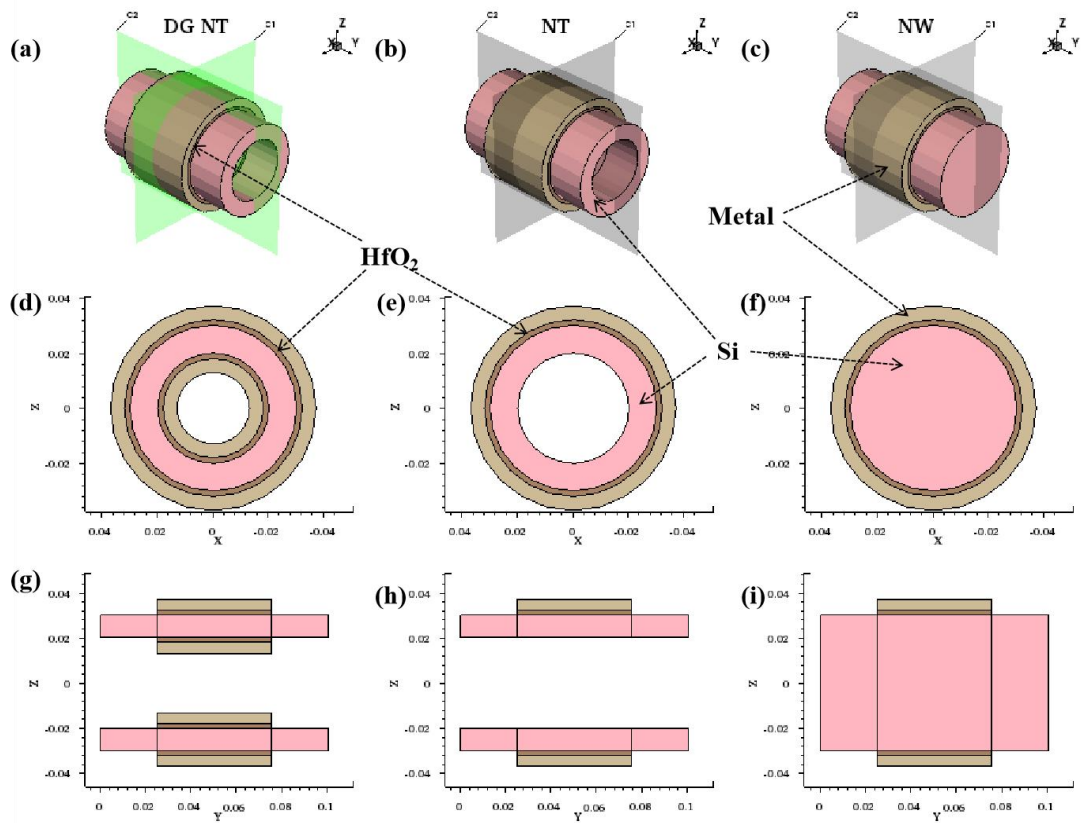

Fig. 1 The 3-D structure of DG NT GAAFET (a), NT GAAFET (b) and NW GAAFET (c) respectively; and their corresponding y-cut profiles perpendicular to (d), (e), (f) and x-cut profiles along (g), (h), (i) the channels.

**Results and Discussions**

Fig. 2 shows the transfer (a), (b) and output (c), (d) characteristics of DG NT GAAFET, NT GAAFET and NW GAAFET under $V_{gs}$ of 0.05V, 1.0V and $V_{ds}$ of 0.5V and 1.0V respectively. According to the simulated results, DG NT GAAFET obtains $I_{on}$ gains of 62% and 57% compared with those of NT GAAFET and NW GAAFET respectively thanks to the enhanced gate to channel coupling brought about by additional inner gate. $I_{off}$ shows 46% and more than 1 order improvements than NT GAAFET and NW GAAFET. SS improves by 11.4% and 66.3% than NT GAAFET and NW GAAFET. $I_{on}/I_{off}$ of DG NT GAAFET is 2 times higher than that of NT GAAFET, and almost 1 order improved than that of NW GAAFET. For another, the $I_{on}$ of NT GAAFET is almost comparable with that of NW GAAFET, as presented in Fig. 2 (c), (d), whereas, $I_{off}$, and SS of NT GAAFET are all tremendously improved than those of NW GAAFET. That is to say, the nanotube channel structure contributes to gate electrostatic control enhancement compared with nanowire channel structure due to thinner body thickness, leading to suppressed SCE and inhibited leakage current, promising for being used in future lower power consumption device. The additional inner gate in DG NT GAAFET both leads to improved SCE and increased $I_{on}$ as well attributed to increased gate number N and widened effective channel length $W_{eff}$.

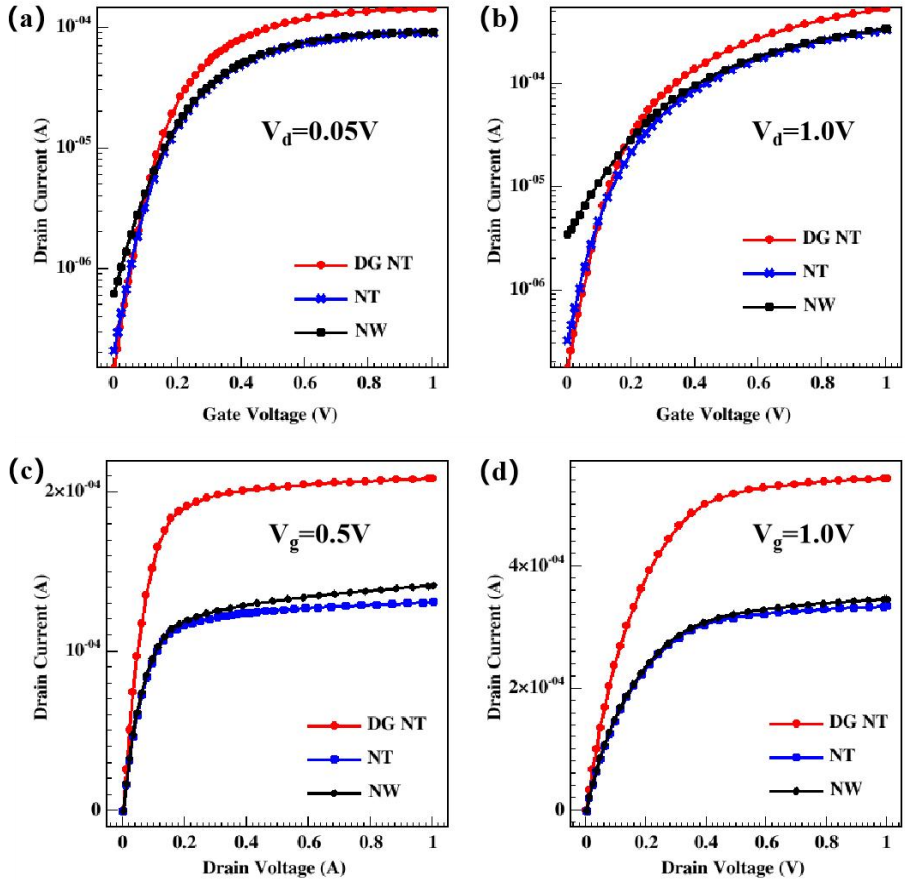

**Fig. 2** $I_d$-$V_g$ transfer characteristics of DG NT GAAFE (the red curves), NT GAAFET (the blue curves), and NW GAAFET (the dark curves) under drain bias of 0.05V (a)

and 1.0V (b) respectively, (c), (d) $I_d$-$V_d$ output curves under $V_{gs}$ of 0.5V and 1V respectively for DG NT GAAFE (the red curve), NT GAAFET (the blue curve), and NW GAAFET (the dark curve).

To clarify the augmentation of $I_{on}$ and improvement of SCE in DG NT GAAFET compared with NT GAAFET and NW GAAFET structures, the electron current density distributions of three kinds of transistor structures along and perpendicular to the the channels have been demonstrated, as shown in Fig. 3. The electron currents centralize on the inner and outer interface regions in DG NT GAAFET, whereas, in NT and NW GAAFET devices, the electron currents concentrate in the surface regions in proximity to the outer gate, as shown in Fig. 3 (b), (c), and (e), (f), which explains the enhanced $I_{on}$ and improved gate electrostatic control ability of DG NT GAAFET compared with NT GAAFET and NW GAAFET. Fig3 (b), (c), and (e), (f) also show that the electron current density of NT GAAFET is more concentrated than that of NW GAAFET attributed to enhanced gate electrostatic control of channel caused by thinner channel as shown in Fig. 4.

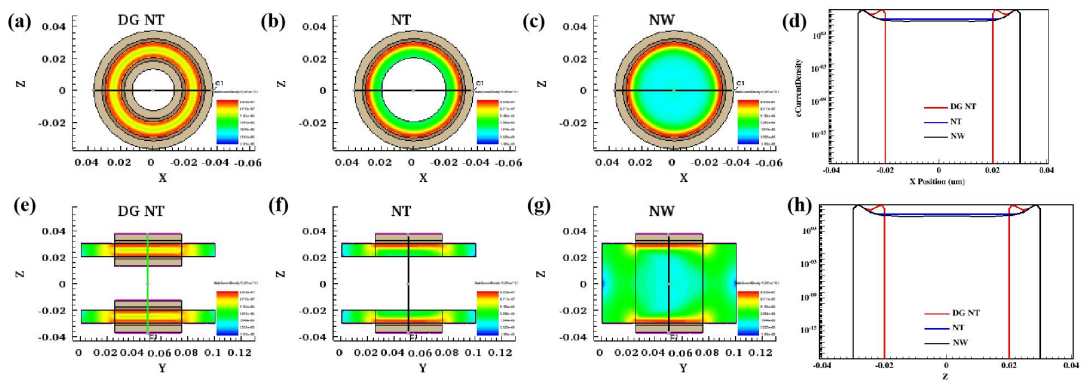

Fig. 3 The electron current density distribution profiles of DG NT GAAFET (a), NT GAAFET, (b), and NW GAAFET (c), perpendicular to the channels direction; (d) The electron current density distribution profiles of DG NT GAAFET, NT GAAFET and NW GAAFET along the cutlines marked in (a), (b), (c); The electron current density distribution profiles of DG NT GAAFET (e), NT GAAFET, (f), and NW GAAFET (g), along the channels direction; (h) The electron current density distribution profiles of DG NT GAAFET, NT GAAFET and NW GAAFET along the cutlines marked in (e), (f), (g).

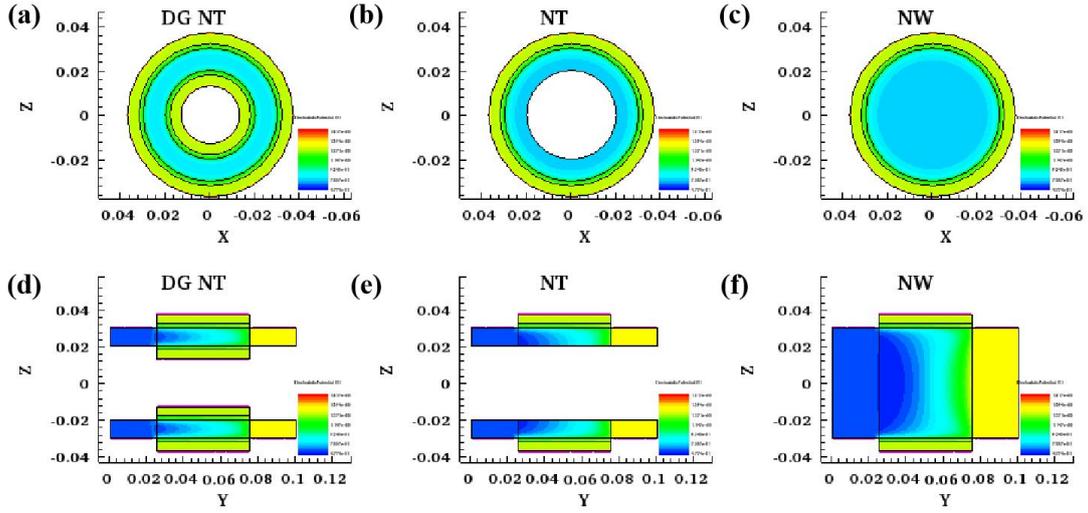

**Fig. 4** Electrostaticpotential distribution profiles of DG NT GAAFET, NT GAAFET and NW GAAFET perpendicular to (a), (b) and (c), and along (d), (e), (f) the channels directions.

Fig. 4 demonstrates the electron static potentials of DG NT GAAFET, NT GAAFET and NW GAAFET structures perpendicular to and along the channels directions respectively. Due to the existence of the inner gate, the electron static potential of DG NT GAAFET is higher than those of NT GAAFET and NW GAAFET, indicating stronger gate to channel coupling ability brought by double-gated structures. For another, the electron static potential in NT GAAFET is more concentrated and stronger than NW GAAFET, the reason lies in that, the channel thickness has been thinned down in NT GAAFET compared with NW GAAFET, thus leading to better gate electronstatic control as well.

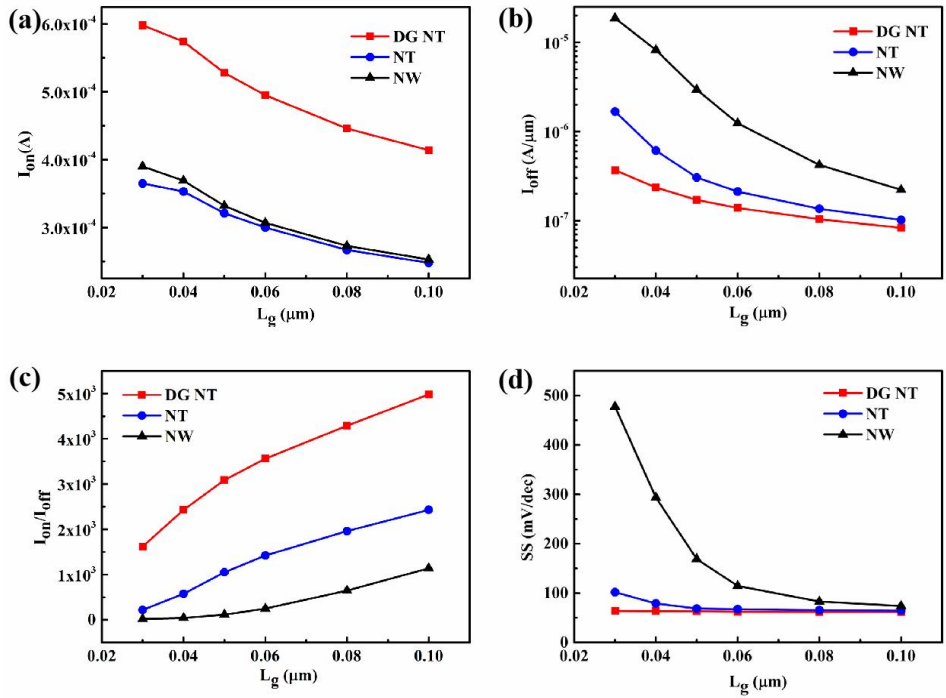

**Fig. 5** (a) $I_{on}$, (b) $I_{off}$, (c) $I_{on}/I_{off}$ and (d) SS vs $L_g$ trends of DG NT GAAFET (the red lines), NT GAAFET (the blue lines) and NW GAAFET (the dark lines) respectively.

The scalability of DG NT GAAFE, NT GAAFET and NW GAAFET have been studied as well, the results are summarized in Fig. 5, where $I_{on}$, $I_{off}$, SS and $I_{on}/I_{off}$ ratio versus $L_g$ trends of DG NT GAAFET, NT GAAFET and NW GAAFET are demonstrated respectively. It can be obtained that $I_{on}$ of DG NT GAAFET is more than 60% and 50% higher than those of NT GAAFET and NW GAAFET at all gate lengths. On the other hand, the $I_{on}$ of NW GAAFET is slightly larger than that of NT GAAFET, and the phenomenon becomes more evident when $L_g$ gets smaller. Fig. 5 (b) and (c) show that DG NT GAAFET is superior than both NT GAAFET and NW GAAFET in $I_{off}$ and $I_{on}/I_{off}$ ratios at all gate lengths simulated, furthermore, NT GAAFET has better $I_{off}$ and $I_{on}/I_{off}$ ratio than NW GAAFET. From Fig. 5(d), it can be concluded that the SS of DG NT GAAFET only changes slightly from 61mV/dec to 66mV/dec, almost unchanged at all gate lengths simulated. Whereas, NT GAAFET and NW GAAFET encounter a relatively serious SS degradation, with SS of NT GAAFET changing from 64mV/dec to 102mV/dec, much better than that of NW GAAFET, which degrades drastically from 74mV/dec to 478mV/dec as $L_g$ changes from 0.1μm to 0.03μm, further supporting the scaling robustness of nanotube channel structure against SCE as $L^g$ scaling down.

**Conclusion**

In conclusion, the gate electrostatic control enhancement in DG NT GAAFET structure as opposed to NT GAAFET and NW GAAFET and the scalability of nanotube channel structure have been investigated with Sentaurus TCAD simulation. $I_{on}$ gains of 62% and 57% have been obtained in DG NT GAAFET compared with NT GAAFET and NW GAAFET devices at $L_g$ of 50nm. In addition, SCE in DG NT GAAFET has been dramatically suppressed thanks to enhanced gate to channel coupling brought about by the additional inner gate, as be indicative of by substantially improved SS, $I_{off}$ and $I_{on}/I_{off}$ ratio. The $I_{on}$ of NT GAAFET is comparable with that of NW GAAFET, whereas, the SS and $I_{off}$ of NT GAAFET improve at all channel length simulated than those of NW GAAFET. Compared with NW GAAFET, both DG NT GAAFET and NT GAAFET showed electrostatic robustness against $L_g$ scaling down, indicating the potential of nanotube channel structure in further device scaling application.

**Acknowledgement**
This work is supported by these funds of 2021Szvup002, JCYJ20210324115812036 JCYJ20200109144612399, JCYJ20200109144601715, IERF202002, IERF202105, IERF202206, JCYJ20220818103408018.

**References**
[1] Naim Hossain Patoary, Jing Xie, Guantong Zhou, Fahad Al Mamun, Mohammed Sayyad, Sefaattin Tongay, Ivan Sanchez Esqueda, Improvements in 2D p-type WSe$_2$ transistors towards ultimate CMOS scaling, Sci. Rep., 13, 3304, 2023.
[2] J. Lakshmi Prasanna, M. Ravi Kumar, Ch. Priyanka, Chella Santhosh, Evaluation of Device Fabrication from FET to CFET: A Review, Journal of Nano and

Electronic Physics, 16, 06030, 2021.

[3] J. Ajayan, D. Nirmal, Shubham Tayal , Sandip Bhattacharya, L. Arivazhagan, A.S. Augustine Fletcher, P. Murugapandiyan, D. Ajitha, Nanosheet field effect transistors-A next generation device to keep Moore's law alive: An intensive study, Microelectronics Journal, 114, 105141, 2021.

[4] Yuan Liu, Xidong Duan, Yu Huang and Xiangfeng Duan, Two-dimensional transistors beyond graphene and TMDCs, Chem. Soc. Rev., 47, 6388, 2018.

[5] Manish Chhowalla, Debdeep Jena, and Hua Zhang, Two-dimensional semiconductors for transistors, Nat Rev Mater 1, 16052, 2016.

[6] Henrique trombini, Gabriel Guterres Marmitt, igor Alencar, Daniel Lorscheitter Baptista, Shay Reboh, frédéric Mazen, Rafael Bortolin Pinheiro, Dario Ferreira Sanchez, carlos Alberto Senna, Bráulio Soares Archanjo, carlos Alberto Achete & Pedro Luis Grande, Unraveling structural and compositional information in 3D finfet electronic devices, Sci. Rep., 9, 11629, 2019.

[7] Ravindra Kumar Maurya & Brinda Bhowmick, Review of FinFET Devices and Perspective on Circuit Design Challenges, Silicon, 14, 5783, 2022.

[8] N. P. Maity, Reshmi Maity, S. Maity, S. Baishya, Comparative analysis of the quantum FinFET and trigate FinFET based on modeling and simulation, Journal of Computational Electronics, 18, 492-499, 2019.

[9] A. Razavieh, P. Zeitzoff, and E.J. Nowak, Challenges and Limitations of CMOS Scaling for FinFET and Beyond Architectures, IEEE Transactions on Nanotechnology, 18, 999-1004, 2019.

[10] Yu-Ru Lin, Yu-Hsien Lin, Yi-Yun Yang, and Yung-Chun Wu, Comprehensive Study of Stacked Nanosheet-Type Channel Based on Junctionless Gate-All-Around Thin-Film Transistors, Journal of the Electron Device Society, 7, 969, 2019.

[11] Doyoung Jang, Dmitry Yakimets, Geert Eneman, Pieter Schuddinck, Marie Garcia Bardon, Praveen Raghavan, Alessio Spessot, Diederik Verkest, and Anda Mocuta, Device Exploration of NanoSheet Transistors for Sub-7-nm Technology Node, IEEE Transactions on Electron Devices, 64, 2707, 2017.

[12] Barraud, S., Previtali, B., Vizioz, C., Hartmann, J. M., Sturm, J., Lassarre, J., ... & Andrieu, F., 7-levels-stacked nanosheet GAA transistors for high performance computing, In 2020 IEEE Symposium on VLSI Technology (pp. 1-2). IEEE.

[13] Jie Gu, Qingzhu Zhang, Zhenhua Wu, Yanna Luo, Lei Cao, Yuwei Cai, Jiaxin Yao, Zhaohao Zhang, Gaobo Xu, Huaxiang Yin, Jun Luo, Wenwu Wang, Narrow Sub-Fin Technique for Suppressing Parasitic-Channel Effect in Stacked Nanosheet Transistors, Journal of the Electron Device Society, 10, 35, 2022.

[14] Yang, X., Li, X., Liu, Z., Sun, Y., Liu, Y., Li, X., & Shi, Y. (2022). Impact of Process Variation on Nanosheet Gate-All-Around Complementary FET (CFET). IEEE Transactions on Electron Devices, 2022, 69, 4029-4036.

[15] Hossain M. Fahad, Casey E. Smith, Jhonathan P. Rojas, and Muhammad M. Hussain, Silicon Nanotube Field Effect Transistor with Core? Shell Gate Stacks for Enhanced High-Performance Operation and Area Scaling Benefits, Nano Lett., 2011, 11, 4393-4399.

[16] Hossain M. Fahad, and Muhammad M. Hussain, High-Performance Silicon Nanotube Tunneling FET for Ultralow-Power Logic Applications, IEEE Transaction on Electron Devices, 60, 1034, 2013.

[17] Daniel Tekleab, Device Performance of Silicon Nanotube Field Effect Transistor, IEEE Electron Device Letters, 35, 506, 2014.

[18] Avtar Singh, Chandan Kumar Pandey, Saurabh Chaudhury, and Chandan Kumar Sarkar, Effect of strain in silicon nanotube FET devices for low power applications, Eur. Phys. J. Appl. Phys. 85, 10101, 2019.

[19] Josephine Anucia. A, Gracia. D & Jackuline Moni D, DC and RF analysis of Vertical 3D p-type Silicon Nanotube FET for low power applications, International Journal of Electronics, 109, 721-734, 2022.